\documentclass[a4paper,11pt]{article}
\usepackage{pos}
\usepackage{hyperref}
\usepackage{url}
\title{Lattice QCD at finite temperature and density}

\author*[a]{Heng-Tong Ding}

\affiliation[a]{Key Laboratory of Quark and Lepton Physics (MOE) and Institute of
Particle Physics, \\ Central China Normal University, Wuhan 430079, China}

\emailAdd{hengtong.ding@ccnu.edu.cn}

\abstract{
I review recent lattice results on strongly interacting matter under extreme conditions, with emphasis on the finite-temperature QCD transition at $\mu_B=0$, its approach toward the chiral limit and the fate of the $U_A(1)$ anomaly, as well as recent constraints on the QCD phase boundary and the possible critical endpoint at $\mu_B>0$. I also discuss selected advances in lattice methods and in QCD thermodynamics under external conditions, in particular strong magnetic fields, isospin chemical potential, rotation, acceleration, and quark spin polarization.
}

\FullConference{The 42nd International Symposium on Lattice Field Theory (LATTICE2025)\\
2-8 November 2025\\
Tata Institute of Fundamental Research, Mumbai, India\\}

\begin{document}
\maketitle
\section{Introduction}
Understanding QCD matter under extreme conditions is a central goal of contemporary nuclear and particle physics, with phenomenology spanning
the early Universe, relativistic heavy-ion collisions, and dense matter in compact stars.
Lattice QCD provides a nonperturbative and systematically improvable approach to compute QCD thermodynamics from first principles~\cite{Ding:2015ona}, and it has reached
high precision at vanishing baryon chemical potential.

These proceedings are based on my plenary talk and are necessarily selective.
Transport coefficients and heavy quarkonium were covered by Dibyendu Bala at this conference~\cite{Bala:2025lat}, while M.~Pepe’s plenary focuses on the high-temperature equation of state~\cite{Pepe:2026wmj}. These topics will therefore not be reviewed here.
I focus on:
(i) the QCD transition at $\mu_B=0$, especially the approach toward the chiral limit and the $U_A(1)$ anomaly;
(ii) recent updates on constraints and estimates of the critical endpoint (CEP) and the QCD phase boundary at $\mu_B>0$;
(iii) selected methodological progress for complex measures relevant to the sign problem; and
(iv) bulk thermodynamics and QCD thermodynamics under external conditions, in particular strong magnetic fields, isospin chemical potential, rotation,
and quark spin polarization.

\section{QCD thermodynamics at $\mu_B=0$}
\subsection{Chiral crossover at the physical point from different discretization schemes}
At the physical quark masses and $\mu_B=0$, lattice QCD has established that the finite-temperature transition is a smooth crossover~\cite{Aoki:2006we}.
Continuum-extrapolated staggered-fermion calculations yield a pseudocritical temperature
$T_{\rm pc}\simeq 156~\mathrm{MeV}$~\cite{HotQCD:2018pds,Borsanyi:2020fev}.
A major recent theme is to cross-check this picture using chirally symmetric lattice fermions, where chiral symmetry and the axial anomaly are treated more faithfully at finite lattice spacing.
A standard and robust probe in these comparisons is the light-quark chiral susceptibility, whose peak defines a pseudocritical temperature at $\mu_B=0$.
In particular, HotQCD simulations with Möbius domain-wall fermions (MDWF) at (near-)physical quark masses find the peak position to be
$T_{\rm pc}=158.7^{+2.6}_{-2.3}\,\mathrm{MeV}$~\cite{Gavai:2024mcj,Bhattacharya:2014ara}, consistent with the staggered continuum value within uncertainties already at comparatively coarse lattices (e.g.\ $N_\tau=8$).

More refined MDWF studies at $N_\tau=12$ and $16$ by JLQCD~\cite{Kanamori:2026sep} show only small cutoff effects and a mild volume dependence of the peak, supporting a crossover rather than a true singularity at the physical point.
Overlap-fermion calculations provide an additional cross-check and likewise find no evidence for a linear-in-volume growth of the peak height~\cite{Kotov:2025lat}.

Conserved-charge fluctuations provide an independent probe of the transition region and are directly connected to experimentally accessible event-by-event cumulants.
They also offer a stringent test of lattice discretization effects, since different fermion formulations can be compared at the level of the same continuum observables.
A particularly nontrivial cross-check is the electric-charge fluctuation, which is pion dominated and therefore potentially sensitive to taste breaking in staggered formulations.
Recent $N_f=2+1$ MDWF results at $N_\tau=12$ and $16$~\cite{Goswami:2025euh,Goswami:2026hit} agree with HISQ determinations~\cite{Bollweg:2021vqf} in the continuum limit, with the MDWF results approaching the continuum from above and the HISQ results from below.
Other combinations show the same pattern of consistency: for example, the charge--strangeness correlator, which is kaon dominated, exhibits similarly good agreement between discretizations.

\subsection{Symmetry restoration and the $U_A(1)$ anomaly}
\label{sec:ua1}

Having shown the established consistent crossover picture at the physical point, we now turn to the
pattern of symmetry restoration around the transition region.
While restoration of $\mathrm{SU}(2)_L\times \mathrm{SU}(2)_R$ is governed by chiral dynamics,
the fate of the $U_A(1)$ anomaly at finite temperature remains pivotal:
it controls the universality class of the two-flavor chiral transition and constrains the Columbia plot near the chiral limit.

\subsubsection{Macroscopic manifestations: correlators, screening masses, and topology}
A practical way to assess symmetry restoration is through (non-)degeneracies of channels related by chiral or axial transformations.
Degeneracy of $\pi$ and $\sigma$ (or vector and axial-vector) probes $\mathrm{SU}(2)_L\times \mathrm{SU}(2)_R$ restoration,
whereas degeneracy of pseudoscalar and scalar in the same isospin channel probes effective $U_A(1)$ restoration.
Equivalently, one often considers integrated correlators (susceptibilities). For instance, with
$\pi^i(x)= i \bar\ell(x)\gamma_5\tau^i \ell(x)$,
$\chi_\pi = \int d^4x\;\langle \pi^i(x)\pi^i(0)\rangle,
~
\chi_\delta = \int d^4x\;\langle \delta^i(x)\delta^i(0)\rangle$,
and $\chi_\pi-\chi_\delta$ (or the disconnected light quark chiral susceptibility or the corresponding screening-mass splitting) is a standard $U_A(1)$ indicator.
In addition, since the axial anomaly is tied to topology, the topological susceptibility
$\chi_t(T)=\langle Q^2\rangle/V$ provides a particularly direct macroscopic manifestation.

\paragraph{Physical-point benchmark from topological susceptibility $\chi_t(T)$.}
At the physical pion mass, $\chi_t(T)$ has recently been determined with several discretizations and provides a clean macroscopic manifestation of the axial anomaly.
Since $\chi_t(T)$ decreases monotonically with temperature, an informative consistency check is how smoothly the finite-$T$ results connect to the $T=0$ baseline as $T$ is lowered.
In HotQCD MDWF simulations at $N_\tau=8$~\cite{Gavai:2024mcj}, the data in the crossover region lie below the HISQ continuum estimate, which may indicate sizable cutoff effects on such coarse lattices.
However, the JLQCD MDWF study at finer lattice spacings ($N_\tau=10,12,16$) further shows that for $T\!\lesssim\!150~\mathrm{MeV}$ results with $N_\tau\!\le\!10$ still deviate visibly from the expected approach to the $T=0$ baseline, so this low-temperature side of the transition requires finer lattices for reliable continuum control of $\chi_t(T)$~\cite{Kanamori:2026sep}.
For $T\gtrsim150~\mathrm{MeV}$ the ordering among the finer-$N_\tau$ data becomes more stable, yet residual finite-$a$ effects can remain non-negligible for topology-related observables,
reflecting in particular the sensitivity to the definition of $Q$ (e.g.\ flow-time window), autocorrelations and sampling of topological sectors, and quark-mass/scale matching.
Overall, the observed offsets between discretizations in the transition region are best understood as different cutoff trajectories rather than as evidence for an inconsistency of the continuum limit.
A complementary continuum study in $N_f=2+1+1$ QCD with Wilson twisted-mass fermions finds $\chi_t$ nonzero up to $T\simeq 600~\mathrm{MeV}$ and that DIGA (Dilute Instanton Gas Approximation)-like behavior sets in only at sufficiently high temperatures (of order a few hundred MeV)~\cite{Kotov:2025ilm}.

\begin{figure}[t]
  \centering
  \includegraphics[width=\linewidth]{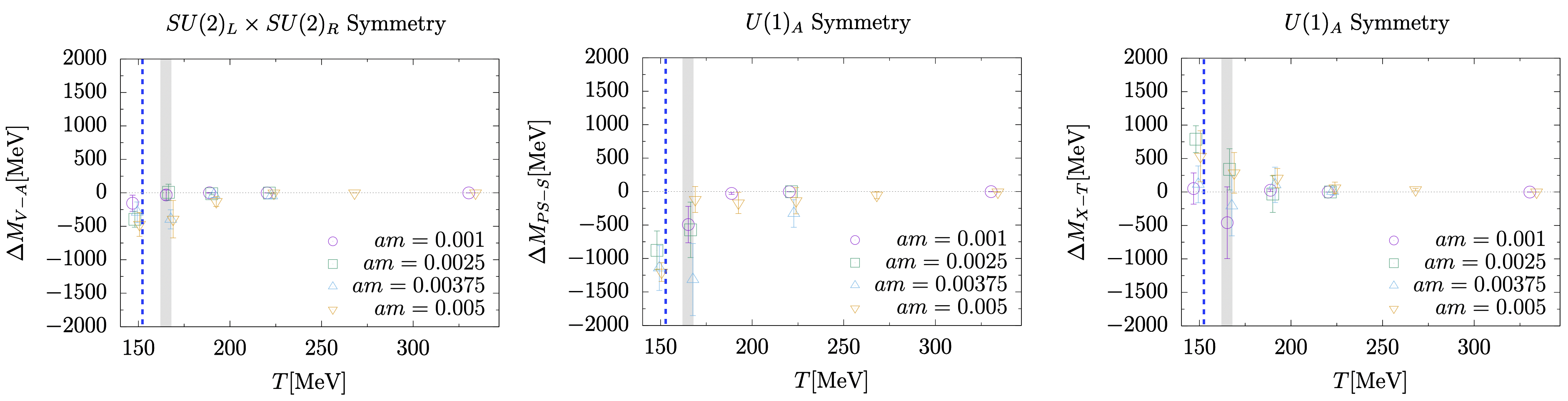}
\caption{Screening-mass restoration patterns toward the chiral limit in $N_f=2$ MDWF simulations~\cite{Aoki:2025mue}. $N_f=2$ MDWF screening masses toward the chiral limit: V--AV (left), PS--S (middle) and X--T splittings (right).}
  \label{fig:screening_nf2}
\end{figure}

\paragraph{Toward the chiral limit: screening masses in $N_f=2$ MDWF.}
While the physical-point results establish a robust baseline, the role of $U_A(1)$ in controlling the chiral-limit transition requires pushing to smaller quark masses.
JLQCD has studied screening masses in two-flavor QCD with MDWF for several quark masses, including one below and several above the physical pion mass~\cite{Aoki:2025mue} (see ~\autoref{fig:screening_nf2}). 
In these simulations the pseudocritical temperature from the chiral susceptibility is $T_{\rm pc}\simeq 165(3)\,\mathrm{MeV}$ (vertical grey band in ~\autoref{fig:screening_nf2}),
while the chiral-limit critical temperature is lower, $T_c\simeq 153~\mathrm{MeV}$ (vertical dashed line)~\cite{Aoki:2024uvl}.
Here $T_c$ is obtained by extracting $T_{\rm pc}(m_l)$ from the peak of the (disconnected) chiral susceptibility at each quark mass and then extrapolating $T_{\rm pc}(m_l)$ toward $m_l\!\to\!0$ using the available mass set (with only one point below the physical mass).
This $N_f=2$ chiral-limit temperature should be contrasted with the continuum-extrapolated chiral-limit critical temperature in $N_f=2+1$ QCD, for which HotQCD scaling analyses and Wilson-fermion studies yield a lower value, $T_c\simeq (132$--$135)\,\mathrm{MeV}$~\cite{HotQCD:2019xnw,Kotov:2021rah}.
The V--AV screening-mass splitting decreases toward smaller quark mass and becomes consistent with zero near $T_c$ at the lightest mass,
signaling $\mathrm{SU}(2)_L\times\mathrm{SU}(2)_R$ restoration.
For $U_A(1)$, the PS--S splitting is more subtle near $T_c$ (channel dependent and at the level of $\sim 1\sigma$),
but at $T\simeq 190~\mathrm{MeV}\approx 1.2\,T_c$ the splittings tend to disappear at the smallest quark mass~\cite{Aoki:2025mue}.

\paragraph{$N_f$ and $M_\pi$ systematics.}
Before turning to microscopic mechanisms, it is important to note that the quantitative size of $U_A(1)$ indicators depends on both the flavor content and the pion mass, and can be further influenced by the lattice setup used to access the relevant correlators.
Using $\mathcal{O}(a)$-improved Wilson fermions on an anisotropic lattice ($\xi=7$), FASTSUM has studied $N_f=2+1$ QCD on $32^3\times N_\tau$ lattices with $N_\tau\in[128,24]$ ($T\simeq 100$--$534~\mathrm{MeV}$) at $M_\pi=378(1)\,\mathrm{MeV}$, and finds PS--S differences tend to vanish at $T\sim 320~\mathrm{MeV}$ with $T_{pc}\sim180~\mathrm{MeV}$ in this setup~\cite{Bignell:2026ybw}.
In contrast, optimal domain-wall simulations probe different flavor content and masses: for $N_f=2+1+1$ at the physical point on $32^3\times N_\tau$ with $N_\tau\in[16,2]$ ($T\simeq 192$--$1540~\mathrm{MeV}$), the V--AV measure becomes small already around $T\simeq 192~\mathrm{MeV}$, while a $U_A(1)$-sensitive splitting remains sizable and disappears only at higher temperatures~\cite{Chiu:2024jyz}.
An additional $N_f=2+1+1+1$ study on $40^3\times N_\tau$ with $N_\tau\in[20,2]$ ($T\simeq 325$--$3252~\mathrm{MeV}$) at $M_\pi\sim 700~\mathrm{MeV}$ shows markedly larger splittings at low $T$~\cite{Chiu:2024bqx}.
Taken together, these results highlight that quantitative statements about “effective” $U_A(1)$ restoration require matched $N_f$, $M_\pi$, and lattice systematics (volume, spacing, and anisotropy) to disentangle genuine physics from setup-dependent effects.

Complementary investigations have also been pursued in other regions of parameter space, including the heavy-quark high-density regime~\cite{Ejiri:2026qjk}, massless many-flavor QCD~\cite{Klinger:2026pbe}, and updated $N_f=3$ MDWF studies with two additional higher-temperature points~\cite{Aoki:2025lat}.

\subsubsection{Dirac eigenvalue spectrum: microscopic mechanism and quantitative diagnostics}
\label{sec:dirac}

A microscopic route to chiral symmetry breaking and the axial anomaly is provided by the Dirac eigenvalue spectrum.
In infinite volume at zero temperature, the Banks--Casher relation connects the chiral condensate to the spectral density at the origin:
$\Sigma \equiv -\lim_{m_\ell\to 0}\lim_{V\to\infty}\langle \bar\psi\psi\rangle
= \pi\,\rho(0)\,$.
At finite temperature, restoration of chiral symmetry implies $\rho(0)\to 0$ in the chiral limit.
The difference of susceptibilities in pseudoscalar and scalar isovector channels can be written in terms of $\rho(\lambda)$ as
\begin{equation}
\chi_\pi-\chi_\delta = \int_0^\infty d\lambda \,\rho(\lambda,m_\ell)\,
\frac{8m_\ell^2}{(\lambda^2+m_\ell^2)^2}\,,
\label{eq:chidiff}
\end{equation}
which makes explicit that near-zero modes dominate $U_A(1)$ breaking in this channel.

\paragraph{Cohen's gap criterion and constraints from analyticity.}
A classic observation due to Cohen is that a strict gap in $\rho(\lambda)$ near $\lambda=0$ is sufficient to yield
$\chi_\pi-\chi_\delta\to 0$ in the chiral limit (effective $U_A(1)$ restoration)~\cite{Cohen:1997hz}.
However, it is not necessary: a nontrivial interplay of quark-mass dependence and infrared spectral structure can maintain
$U_A(1)$ breaking even when chiral symmetry is restored.

A particularly influential analytic argument is due to Aoki--Fukaya--Taniguchi~\cite{Aoki:2012yj}.
Assuming (i) restoration of $\mathrm{SU}(2)_L\times\mathrm{SU}(2)_R$, (ii) analyticity in $m^2_\ell$ for $m$-independent observables in the chirally symmetric phase, and (iii) a smooth analytic expansion of the thermodynamic-limit eigenvalue density near $\lambda=0$ in the positive-eigenvalue convention, they derive strong constraints on the infrared spectrum. In particular, $\rho(0)=0$, $\rho^\prime(0)=O(m_\ell^2)$ and $\rho^{\prime\prime}(0)=O(m_\ell^2)$ at small nonzero $m_\ell$, so that in the chiral limit the leading nonvanishing term is $\rho(\lambda)\sim\lambda^3$ for $\lambda\ge0$, equivalently $\rho(\lambda)\sim |\lambda|^3$ if expressed in terms of the signed eigenvalue.
Within this analytic framework the $U_A(1)$ anomaly becomes invisible in the scalar/pseudoscalar susceptibilities in the chiral limit,
and correspondingly $U_A(1)$-breaking effects do not show up in scalar/pseudoscalar correlators at least up to six-point functions under those assumptions~\cite{Aoki:2012yj,Kaczmarek:2021ser}. Recent works have revisited these assumptions and discussed singular near-zero-mode
scenarios, including instanton-gas-inspired ones, that can keep $U_A(1)$ broken
in the $N_f=2$ chiral limit~\cite{Kovacs:2023vzi,Giordano:2024jnc,Giordano:2025shr,Giordano:2025fcr,Giordano:2025vbb}.
This motivates looking for more differential diagnostics of the infrared spectrum and, in particular, its quark-mass dependence.

\paragraph{A useful parametrization of $\rho(\lambda)$.} A convenient way to organize possible infrared behaviors is to consider an ansatz of the form~\cite{HotQCD:2012vvd,Buchoff:2013nra,Heng-Tong:2023arw}, 
$\rho(\lambda,m_\ell)=c_0+c_1\lambda+c_2\, m_\ell^2\delta(\lambda)+c_3\, m_\ell + c_4\, m_\ell^2+\mathcal{O}(\lambda,m_\ell)\,,$
which can realize restored $\mathrm{SU}(2)_L\times \mathrm{SU}(2)_R$ but broken $U_A(1)$ in the $m_\ell\to 0$ limit through a near-zero-mode contribution.

In this language, $c_0$ and $c_1$ would break both $\mathrm{SU}(2)_L\times\mathrm{SU}(2)_R$ and $U_A(1)$ and should be absent once chiral symmetry is restored;
the $c_2$ term represents a near-zero-mode contribution (consistent with an infrared peak that sharpens toward a $\delta$-like structure) that can keep $U_A(1)$ broken while allowing $\rho(0)\to 0$;
$c_3$ provides another $U_A(1)$-breaking contribution but is incompatible with strict analyticity in $m_\ell^2$;
and $c_4$ is subleading, contributing only terms that vanish as $m_\ell\to 0$ in two-point $U_A(1)$ measures.
In practice, disentangling a would-be $\delta(\lambda)$-like near-zero-mode component is challenging:
at finite volume and finite lattice spacing the peak is smeared, and its contribution can be largely
degenerate with smooth, analytic $m_\ell$-dependent terms in $\rho(\lambda,m_\ell)$ (e.g.\ $\propto m_\ell,\,m_\ell^2$)
that likewise vanish as $m_\ell\to 0$~\cite{HotQCD:2012vvd,Buchoff:2013nra,Heng-Tong:2023arw}.
One possible way out is to exploit the distinct quark-mass dependence of the various contributions: in particular, an $m_\ell^2\delta(\lambda)$ term leaves a characteristic imprint on quark-mass derivatives of $\rho(\lambda,m_\ell)$.
Naively, accessing such derivatives would require generating additional ensembles with infinitesimally shifted quark masses, which is computationally costly.

\paragraph{Mass-derivative method and evidence for $\rho\propto m_\ell^2$ with a $\delta$-like IR peak.}
A methodology has been developed to evaluate the quark-mass dependence of the Dirac spectrum directly through mass derivatives
$\partial^n\rho(\lambda,m_\ell)/\partial m_\ell^n$, which can be computed at a single simulated quark mass and therefore do not require generating additional ensembles at infinitesimally shifted quark masses~\cite{Ding:2020xlj}.

\begin{figure}[!htp]
  \centering
  \includegraphics[width=0.45\linewidth]{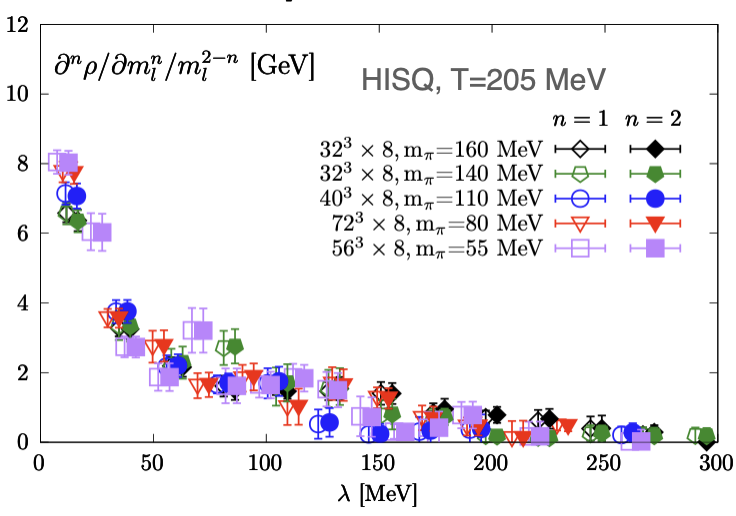}~
  \includegraphics[width=0.45\linewidth]{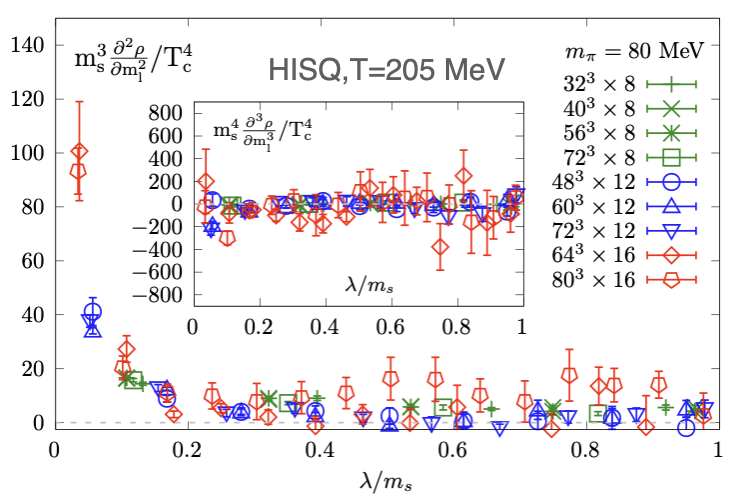}
\caption{Quark-mass derivatives of the Dirac eigenvalue spectrum $\rho(\lambda,m_\ell)$ in $N_f=2+1$ QCD at $T\simeq205~\mathrm{MeV}$.
The near-coincidence of $(\partial\rho/\partial m_\ell)/m_\ell$ with $\partial^2\rho/\partial m_\ell^2$, together with $\partial^3\rho/\partial m_\ell^3\to 0$ within uncertainties, indicates $\rho(\lambda\!\to\!0,m_\ell)\propto m_\ell^2$ in the infrared.
The infrared peak sharpens as the lattice spacing is reduced, suggesting an emerging $\delta(\lambda)$-like near-zero-mode contribution. Plots are taken from Ref.~\cite{Ding:2020xlj}.}
  \label{fig:drho_mass_deriv}
\end{figure}

In $N_f=2+1$ QCD at $T\simeq 205~\mathrm{MeV}$, the derivatives for $n=1,2,3$ exhibit a striking pattern (see~\autoref{fig:drho_mass_deriv}):
$(\partial\rho/\partial m_\ell)/m_\ell$ nearly coincides with $\partial^2\rho/\partial m_\ell^2$, while the third derivative is consistent with zero,
implying $\rho(\lambda\!\to\!0,m_\ell)\propto m_\ell^2$ in the relevant infrared regime~\cite{Ding:2020xlj}.
In addition, the near-zero-mode structure shows a pronounced infrared peak whose width decreases as the lattice spacing is reduced; 
this sharpening toward the continuum suggests an emerging $\delta(\lambda)$-like contribution and supports a DIGA-inspired interpretation of the near-zero-mode sector in this temperature regime.
A direct consequence is that, at $T\simeq 205~\mathrm{MeV}$, the near-zero-mode sector yields a nonvanishing contribution to $U_A(1)$-sensitive observables such as $\chi_\pi-\chi_\delta$
(and related disconnected quantities) in the combined chiral and continuum limits (see Fig. 3 in Ref.~\cite{Ding:2020xlj}). Since the corresponding kernel in Eq.~\eqref{eq:chidiff} weights the infrared region more strongly as $T$ is lowered and the anomaly effects are expected to become more pronounced toward the transition,
this microscopic mechanism naturally implies that $U_A(1)$ breaking should remain visible---and potentially even stronger---as one approaches $T_c$ from above.

Taken together, the persistence of a near-zero-mode contribution to $U_A(1)$-sensitive observables down to the transition region supports the standard scenario that $U_A(1)$ remains effectively broken at $T_c$ in the $m_\ell\!\to\!0$ limit.
In that case, the chiral phase transition of $N_f=2+1$ QCD at physical $m_s$ and vanishing light-quark mass is expected to be second order in the $\mathrm{O}(4)$ universality class (up to the usual $\mathrm{O}(2)$ remnant at finite lattice spacing for staggered formulations).

\begin{figure}[!htp]
  \centering
  \includegraphics[width=0.45\linewidth]{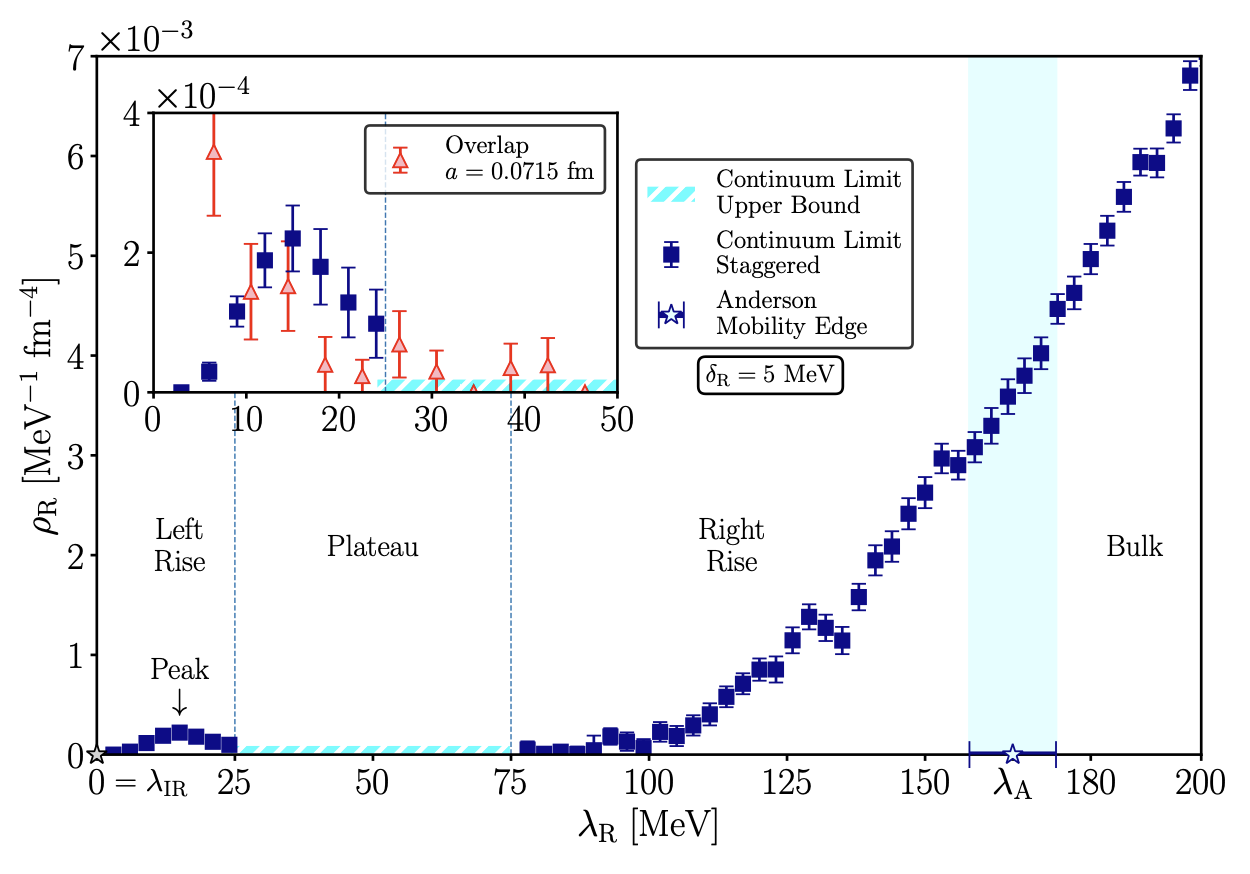}
  \includegraphics[width=0.45\linewidth]{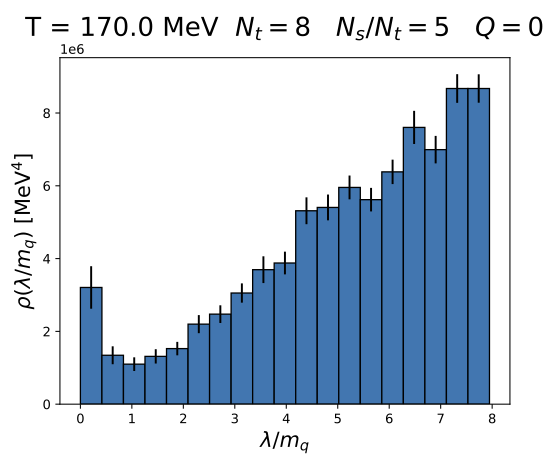}
\caption{Dirac eigenvalue spectrum $\rho(\lambda,m_\ell)$ in $N_f=2+1$ QCD. Left: Obtained using stout staggered fermion at T=230 MeV~\cite{Alexandru:2024tel}. Right: Obtained using overlap fermions at T=170 MeV~\cite{Kotov:2025lat}.}
  \label{fig:rho_peak}
\end{figure}

\paragraph{Infrared peak structures and eigenmode geometry.}
Finally, peak/bump structures in the infrared spectrum have been observed independently as shown in~\autoref{fig:rho_peak}, 
e.g.\ as a bump at $T\simeq 230~\mathrm{MeV}$ with staggered fermions at the physical point~\cite{Alexandru:2024tel},
consistent with overlap results, and as a pronounced IR peak in large-volume overlap studies at fixed topology~\cite{Fodor:2025mqi,Kotov:2025lat}.
Beyond $\rho(\lambda)$ itself, recent analyses of eigenmodes find a characteristic distribution of fractal dimensions
(with median around $2.4$--$2.6$ in the transition region) and explore thermalization diagnostics via Lyapunov exponents~\cite{Pandey:2024goi,Guin:2025lpy}.
These developments reinforce the Dirac-spectrum perspective as a sharp and mechanism-oriented complement to macroscopic probes.

\subsection{Universality, scaling, and a generalized Banks--Casher relation}
\label{sec:scaling}
The approach toward the chiral limit is naturally phrased in terms of universality and scaling.
Near a critical line, the free energy density can be decomposed into singular and regular parts,
$f(m_\ell,T)=f_s(z)+f_{\rm reg}(m_\ell,T)$ with $z=\frac{t}{h^{1/\beta\delta}}$,
$t=\frac{T-T_c}{T_c}$ and $h\propto m_\ell$.
The order parameter and its susceptibilities have scaling forms,
\begin{equation}
M = h^{1/\delta} f_1(z)+M_{\rm reg},\qquad
\chi_M = \frac{1}{h_0} h^{1/\delta-1} f_2(z)+\chi_{M,{\rm reg}},
\end{equation}
and similarly for higher derivatives with respect to $m_\ell$.
A practical issue is to control the regular contributions $M_{\rm reg},\chi_{M,{\rm reg}},\ldots$.
Two useful strategies are:
(i) construct combinations of $M$ and $\chi_M$ that suppress leading regular terms, and
(ii) use higher-order susceptibilities/cumulants, which are increasingly dominated by the singular part.

\paragraph{Suppressing regular terms: improved order parameter.}
A concrete example is the improved order parameter $\tilde M$ proposed in recent work, schematically of the form
$\tilde M = M_\ell - H\,\chi_\ell$, which suppresses the leading regular contribution in $H$~\cite{Mitra:2025aeu,Ding:2024sux,Mitra:2026bbg,Mitra:2024mke,Mitra:2025hsk}.
In this construction, universal scaling behavior becomes visible with controlled finite-volume effects.
Such improved observables are particularly useful when using staggered ensembles at several pion masses down to $m_\pi \sim 55$ MeV.

\paragraph{A generalized Banks--Casher relation for higher cumulants.}
A major development is a framework that relates higher-order cumulants of the chiral condensate to correlations of Dirac eigenvalues,
generalizing Banks--Casher beyond the first cumulant~\cite{Ding:2023oxy}.
Denote the $n$-th order cumulant of $\bar\psi\psi$ by $\mathbb{K}_n(\bar\psi\psi)$.
One can express $\mathbb{K}_n(\bar\psi\psi)=\int_0^\infty P_n(\lambda)\,d\lambda$,
where $P_n(\lambda)$ is built from connected correlations of a regulated spectral density,
$\rho_U(\lambda)=\sum_j \delta(\lambda-\lambda_j),~
P_U(\lambda;\epsilon)=\frac{4\epsilon\,\rho_U(\lambda)}{\lambda^2+\epsilon^2}.$

The generalized Banks--Casher relation states that, in the chiral limit~\cite{Ding:2023oxy},
\begin{equation}
\lim_{m_\ell\to 0} \mathbb{K}_n(\bar\psi\psi) = (2\pi)^n\,\mathbb{K}_n[\rho_U(0)].
\end{equation}
Two practical advantages stand out:
(i) only the infrared part of the Dirac spectrum is needed, and
(ii) the construction can avoid additional fermion inversions beyond those required for the spectral information.

\begin{figure}[t]
  \centering
  \includegraphics[width=0.4\linewidth]{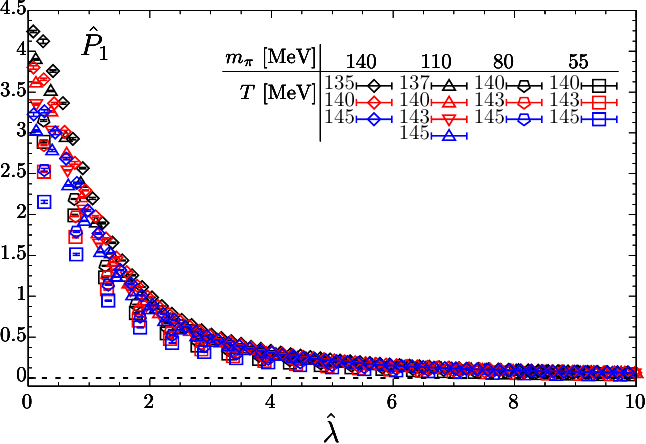}~~~
  \includegraphics[width=0.4\linewidth]{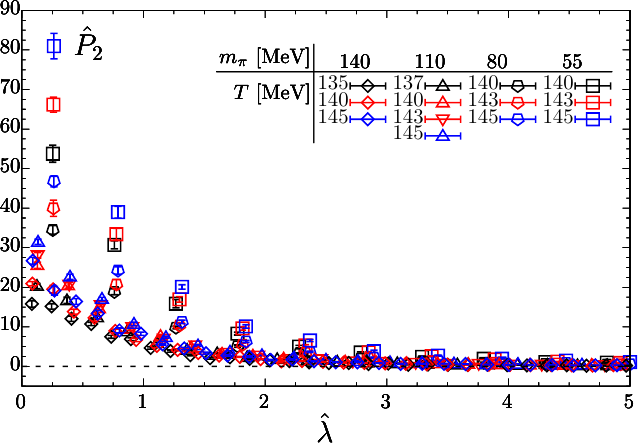}\\
    \includegraphics[width=0.4\linewidth]{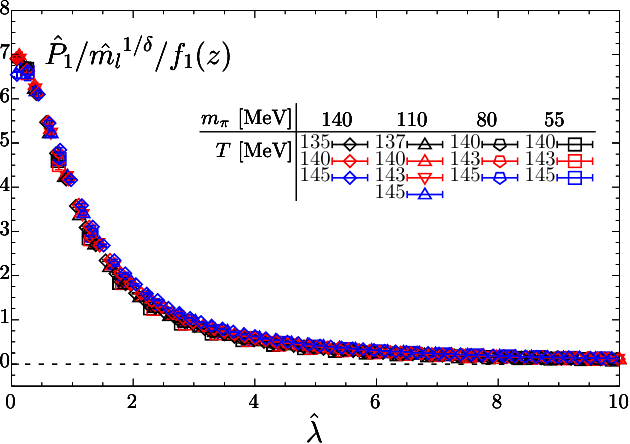}~~~
  \includegraphics[width=0.4\linewidth]{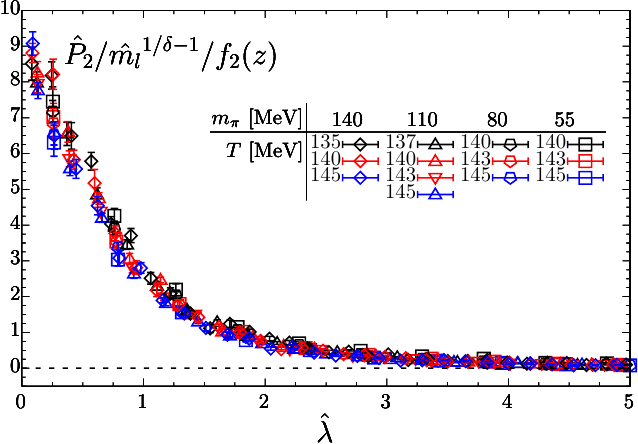}
\caption{From left to right: $P_1$, $P_2$ (top), and rescaled $P_1$ and $P_2$ (bottom) with corresponding macroscopic critical behavior. Figures are taken from~\cite{Ding:2023oxy}.}
  \label{fig:Pn}
\end{figure}

Near criticality, scaling predicts $\mathbb{K}_n(\bar\psi\psi)\sim m_\ell^{1/\delta-n+1}\,f_n(z)$,
providing a direct bridge between microscopic spectral correlations and macroscopic critical behavior. Following the presentation in the Lattice 2025 plenary talk on which this proceedings contribution is based, this connection is visualized through the analogy to ``critical opalescence'' in ordinary fluids: large fluctuations near the critical point are
encoded in enhanced cumulants, and here those cumulants are tied to correlations of near-zero Dirac eigenvalues as shown in~\autoref{fig:Pn}.

\section{Finite density: CEP status, phase boundary, and low-$T$ strategies}
\label{sec:finite_density}
At nonzero baryon chemical potential, lattice QCD is hampered by the sign problem: for real $\mu_B$ the fermion determinant becomes complex, preventing
straightforward importance sampling.
As a result, most first-principles information at $\mu_B\neq 0$ is obtained via expansions around $\mu_B=0$ (Taylor expansion),
analytic continuation from imaginary chemical potential, and more recently, analyses based on Lee-Yang edge singularities using simulations at imaginary chemical potential, as well as other complementary approaches to complex measures.

\paragraph{Upper bound on $T_{\rm CEP}$ from the chiral critical line.}
A robust lattice result is that the chiral transition temperature in the chiral limit decreases with increasing $\mu_B$.
Combining universality and scaling around the chiral critical line with the existence of a tri-critical point separating the
second-order chiral line from a first-order region yields an upper bound on $T_{\rm CEP}$.
In HotQCD analyses one parameterizes the chiral critical line as
$T_c^0(\mu_B) = T_c^0(0)\left(1-\kappa_2\,\hat\mu_B^2+\mathcal{O}(\hat\mu_B^4)\right),$
with continuum and chiral extrapolated values
$T_c^0(0)=132^{+3}_{-6}$ MeV~\cite{HotQCD:2019xnw}, and $\kappa_2\simeq 0.015(1)$ obtained at $N_\tau=8$~\cite{Ding:2024sux} and $\hat{\mu}_B\equiv \mu_B/T$.
For a representative $\hat\mu_B=2$, this implies $T_c^0(\hat\mu_B=2)\lesssim 125$ MeV, providing a conservative upper bound on the CEP temperature
within the assumptions of the argument.

\paragraph{Predictions and updated constraints on the CEP.}
Beyond general bounds, a number of lattice-informed estimates of the CEP location have appeared in recent years~\cite{Schmidt:2025ppy}.
A useful overview is based on the compilation plot shown in~\cite{Lysenko:2024hqp}, which indicates that existing CEP estimates broadly populate
$T_{\rm CEP}\sim 80$--$120~\mathrm{MeV}$ and $\mu_{B,{\rm CEP}}\sim 400$--$650~\mathrm{MeV}$, depending on the methodology and assumptions.
Representative examples include:
(i) a Lee--Yang-edge (LYE) extrapolation from simulations at imaginary $\mu_B$ by the Bielefeld--Parma group (on $N_\tau=6$),
\(
(T_{\rm CEP},\mu_{B,{\rm CEP}})=(105^{+8}_{-18},\,422^{+80}_{-35})~\mathrm{MeV}
\)~\cite{Clarke:2024ugt};
(ii) an LYE-based analysis constrained by HotQCD Taylor coefficients up to $\mathcal{O}(\mu_B^8)$ (on $N_\tau=8$),
yielding an illustrative estimate
\(
(T_{\rm CEP},\mu_{B,{\rm CEP}})\approx(100,\,580)~\mathrm{MeV}
\)~\cite{Basar:2023nkp};
(iii) a constant-entropy construction using lattice inputs for the entropy density and $\mu_B$-dependence at $\mu_B=0$ (with $\mu_Q=\mu_S=0$),
reporting
\(
(T_{\rm CEP},\mu_{B,{\rm CEP}})=(114.3\pm6.9,\,602.1\pm62.1)~\mathrm{MeV}
\)~\cite{Shah:2024img}.
Functional approaches constrained by lattice information give compatible ranges with
\(
(T_{\rm CEP},\mu_{B,{\rm CEP}})\approx(105\text{--}115,\,600\text{--}650)~\mathrm{MeV}
\)~\cite{Fischer:2026uni}.
Overall, these estimates tend to lie at relatively low $T_{\rm CEP}$, consistent with the lattice upper bound $T_{\rm CEP}\lesssim 125~\mathrm{MeV}$ inferred from the chiral-limit curvature analysis discussed earlier.

Updated results from Wuppertal--Budapest provide significantly strengthened constraints on the CEP location based on first-principles lattice input.
In particular, building on the constant-entropy strategy, they combine a new high-precision, continuum-extrapolated $\mu_B=0$ equation of state (using $N_\tau=8,10,12,16$)
with new simulations at imaginary chemical potential under strangeness neutrality, and scan constant-entropy contours continued from imaginary to real $\mu_B$.
With the current precision they exclude, at the $2\sigma$ level, the existence of a CEP for $\mu_B<450~\mathrm{MeV}$~\cite{Borsanyi:2025dyp}.
A complementary Lee--Yang-zero analysis of high-statistics data on $16^3\times 8$ lattices finds a best-fit
$T_{\rm CEP}=55.8^{+6.3}_{-7.9}~\mathrm{MeV}$ and implies an $84\%$ probability that the CEP either lies at $T_{\rm CEP}<103~\mathrm{MeV}$ or does not exist~\cite{Adam:2025phc}.
An important caveat applies to this Lee--Yang-zero inference: it relies on modeling/extrapolating the temperature dependence of the Lee--Yang zeros beyond the temperature window directly constrained by the available lattice input (typically $T\gtrsim120$--$130~\mathrm{MeV}$), motivating more direct first-principles information at lower temperatures.

\begin{figure}[t]
  \centering
  \includegraphics[width=0.33\linewidth]{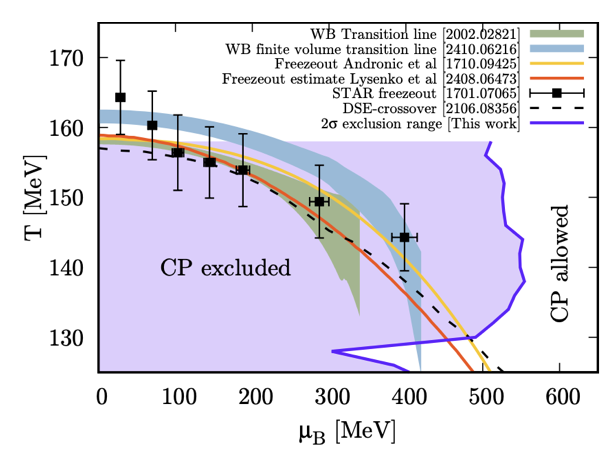}
  \includegraphics[width=0.33\linewidth]{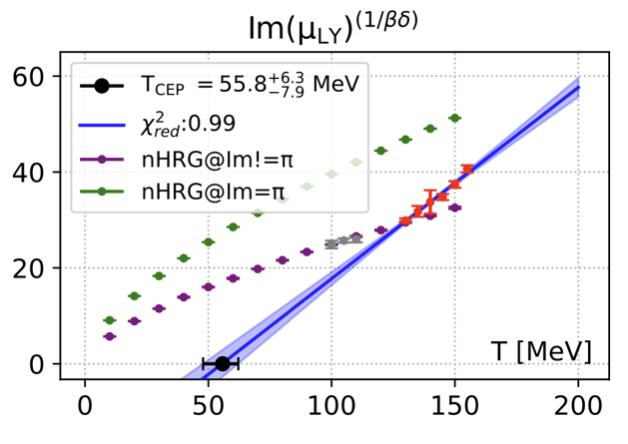}
    \includegraphics[width=0.31\linewidth]{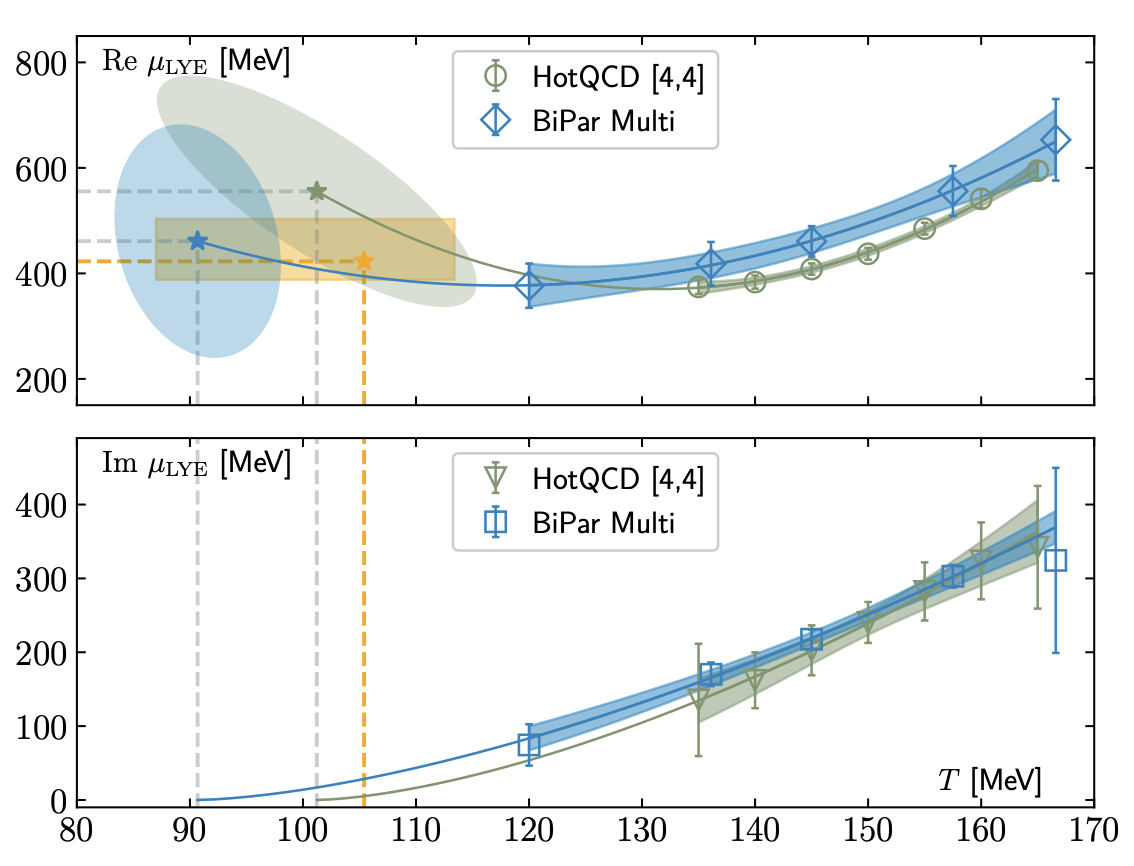}
\caption{Status of lattice constraints and recent estimates for the CEP location.
Left: exclusion region from Wuppertal--Budapest, including the bound of no CEP for $\mu_B<450~\mathrm{MeV}$ at $2\sigma$~\cite{Borsanyi:2025dyp}.
Middle: Wuppertal--Budapest Lee--Yang-edge analysis, yielding an $84\%$ probability for that the CEP lies at $T_{\rm CEP}<103~\mathrm{MeV}$ or does not exist~\cite{Adam:2025phc}.
Right: Parma--Bielefeld Lee--Yang-edge extrapolation from imaginary-$\mu_B$ simulations~\cite{Clarke:2024ugt}.}
  \label{fig:cep_status}
\end{figure}

\paragraph{QCD phase boundary at nonzero $\mu_B$.}
Along the crossover boundary, two complementary lattice strategies have recently been pursued: strangeness-neutral proxy contours in staggered formulations, and hadronic correlator probes in Wilson-fermion calculations.

One strategy pursued by Wuppertal--Budapest focuses on strangeness-dominated proxies under strangeness neutrality. In Ref.~\cite{Borsanyi:2025kiv}, it is observed (from imaginary-$\mu_B$ data) that the chiral susceptibility peak occurs at apparently universal values
$\chi_2^S \simeq 0.3$ and $\mu_S/\mu_B\simeq 0.25$, nearly independent of $\mu_B$ over a wide imaginary range (up to $\mu_B/T \simeq i\,2.75$),
motivating the use of constant-$\chi_2^S$ or constant-$\mu_S/\mu_B$ contours as practical proxies for the crossover line.
Using a two-dimensional expansion in $(\mu_B,\mu_S)$ with 4HEX-improved staggered fermions and aspect ratio $LT=2$ on $16^3\!\times\!8$, $20^3\!\times\!10$ and $24^3\!\times\!12$ lattices,
continuum limits for these proxy contours are obtained up to $\mu_B\simeq400~\mathrm{MeV}$ (with small finite-volume effects assessed by comparing $LT=2$ and $LT=3$),
and on the $16^3\!\times\!8$ lattice the extrapolation can be pushed beyond $\mu_B\simeq500~\mathrm{MeV}$ using very high-order information~\cite{Borsanyi:2025kiv}.
Interestingly, the proxy contours show excellent agreement with HRG expectations and suggest that a deviation from the chemical freeze-out curve may set in only around $\mu_B\sim500~\mathrm{MeV}$.

Wilson-fermion approaches provide a qualitatively different handle based on hadronic correlators.
Ref.~\cite{Smecca:2024gpu} determines $T_{\rm pc}(\mu_B)$ from the condition of vector--axial-vector degeneracy, using $\mathcal{O}(a)$-improved Wilson fermions on anisotropic FASTSUM ensembles and the $\mathcal{O}(\mu_B^2)$ response of mesonic correlators.
Fitting $T_{\rm pc}(\mu_B)/T_{\rm pc}(0)=1-\kappa(\mu_B/T_{\rm pc}(0))^2+\cdots$ yields $\kappa=0.0131(23)(23)$ on the heavier-mass ensemble ($M_\pi\simeq391~\mathrm{MeV}$),
while the lighter ensemble ($M_\pi\simeq239~\mathrm{MeV}$) gives $\kappa=0.034(14)$ with a larger statistical uncertainty~\cite{Smecca:2024gpu}.
Although these results are not yet continuum-extrapolated and remain away from the physical point, they are consistent with the curvature obtained from thermodynamic observables in staggered formulations and illustrate the potential of hadronic probes for constraining the phase boundary.

\begin{figure}[!htp]
  \centering
  \includegraphics[width=0.48\linewidth]{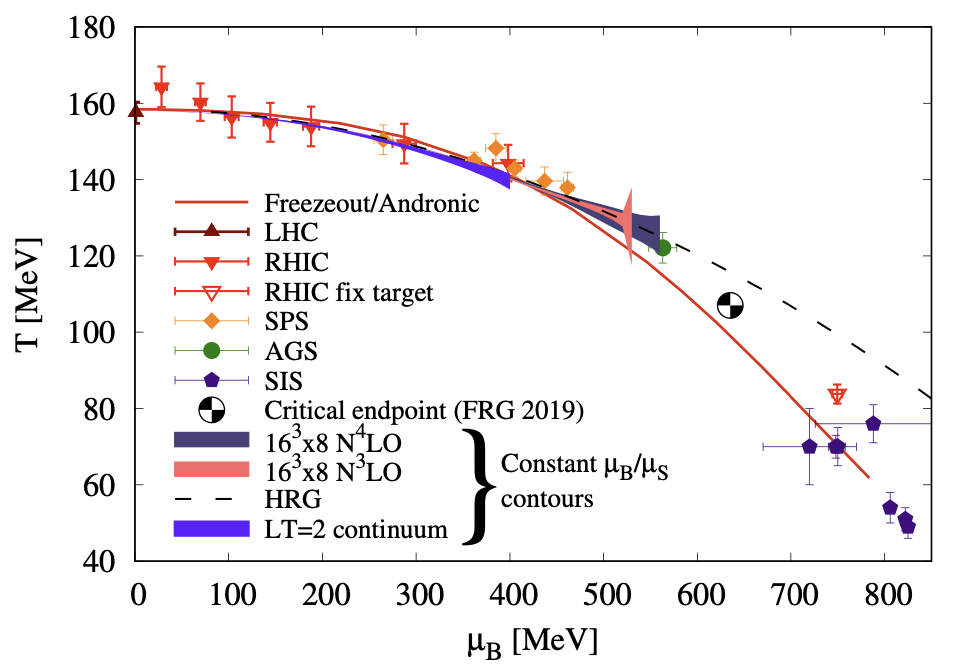}
    \includegraphics[width=0.48\linewidth]{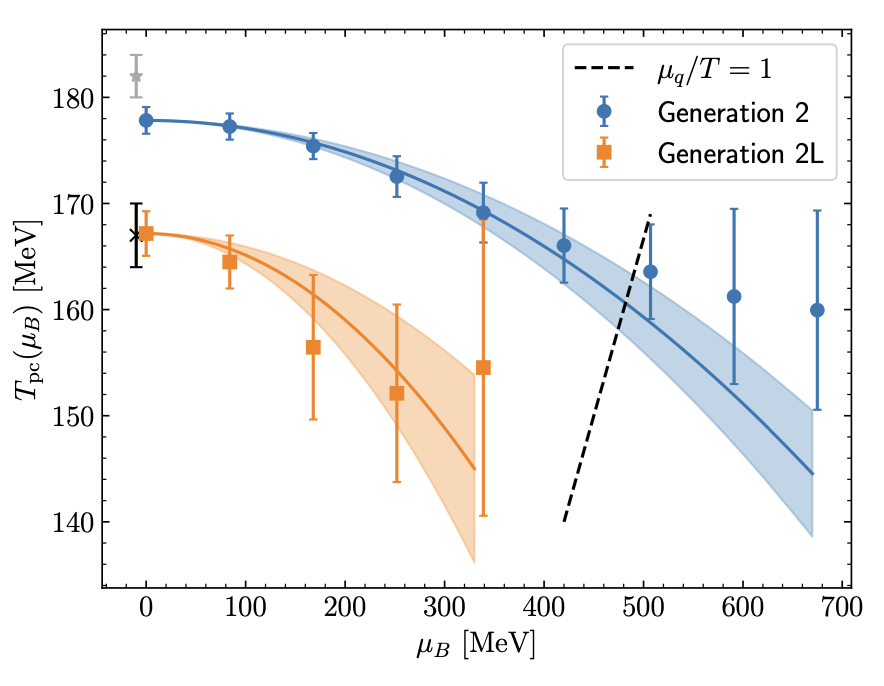}
\caption{Recent lattice constraints on the crossover boundary $T_{\rm pc}(\mu_B)$.
Left: Wuppertal--Budapest proxy contours under strangeness neutrality (e.g.\ constant $\chi_2^S$ / constant $\mu_S/\mu_B$) and their continuum extrapolation~\cite{Borsanyi:2025kiv}.
Right: FASTSUM/Wilson-fermion determination of the curvature of $T_{\rm pc}(\mu_B)$ from vector--axial-vector degeneracy~\cite{Smecca:2024gpu}.}
  \label{fig:phase_boundary}
\end{figure}

\paragraph{Selected tools for complex measures.} 
Methodological progress remains important for extending first-principles reach into regimes where the sign problem is severe.
Here I briefly highlight several representative directions, including diagnostics for complex Langevin (CL) simulations and recent ideas based on diffusion models and worldvolume HMC.

Complex Langevin defines a stochastic process in a complexified configuration space.
Since there is no general proof of correctness for generic complex actions, robust diagnostics are essential.
A widely used criterion is the falloff of the distribution of the drift term:
exponential falloff is compatible with correctness, whereas slow (e.g.\ power-like) tails can indicate incorrect convergence~\cite{Nagata:2016vkn,Ito:2020mys}. 
However, passing such criteria (and even having small boundary terms) is not, by itself, sufficient: CL can still converge incorrectly due to contributions from unwanted
integration cycles in the complexified manifold, motivating kernel control and stricter correctness conditions~\cite{Mandl:2025ins}.
For this reason it is important to employ complementary diagnostics, including configurational-temperature tests and systematic cross-checks across parameters~\cite{Joseph:2025fcd,Joseph:2025xbn,Joseph:2026xti,Longia:2026doi}.
In practice, the drift-distribution criterion remains the workhorse and is now being applied in increasingly realistic settings; for example, CL simulations with four-flavor staggered fermions on $24^3\times 12$ lattices use drift-histogram diagnostics to map out a validity window in $(T,\mu)$ and to identify its breakdown near the transition, providing useful benchmarks~\cite{Tsutsui:2025jez}.

An emerging theme is to combine ideas from machine learning with field-theory algorithms to improve sampling in complex-weight systems. Diffusion-based generative models have been explored first in simple settings such as scalar field theories, and more recently in non-Abelian lattice gauge theories including two-dimensional SU(2) systems~\cite{Wang:2023exq,Aarts:2026zzr,Alharazin:2026lcb}.
In parallel, the worldvolume hybrid Monte Carlo algorithm provides a way to sample complexified manifolds related to Lefschetz thimbles.
Both directions are under active development; their conceptual unification---controlled sampling on complexified manifolds combined with expressive probabilistic
models---is promising for future applications~\cite{Fukuma:2020fez,Fukuma:2025gya}.

\section{Bulk thermodynamics and lattice--experiment interface}
\label{sec:bulk}

The phase boundary and the possible location of a critical endpoint are ultimately questions about the thermodynamic response of QCD matter. This naturally leads to bulk observables: the equation of state and its derivatives, especially fluctuations and correlations of conserved charges. These quantities are both first-principles computable on the lattice and experimentally accessible (as event-by-event cumulants), providing a concrete lattice--experiment interface once acceptance cuts, efficiencies, and non-equilibrium effects are accounted for.

\begin{figure}[!htp]
  \centering
    \includegraphics[width=0.31\linewidth]{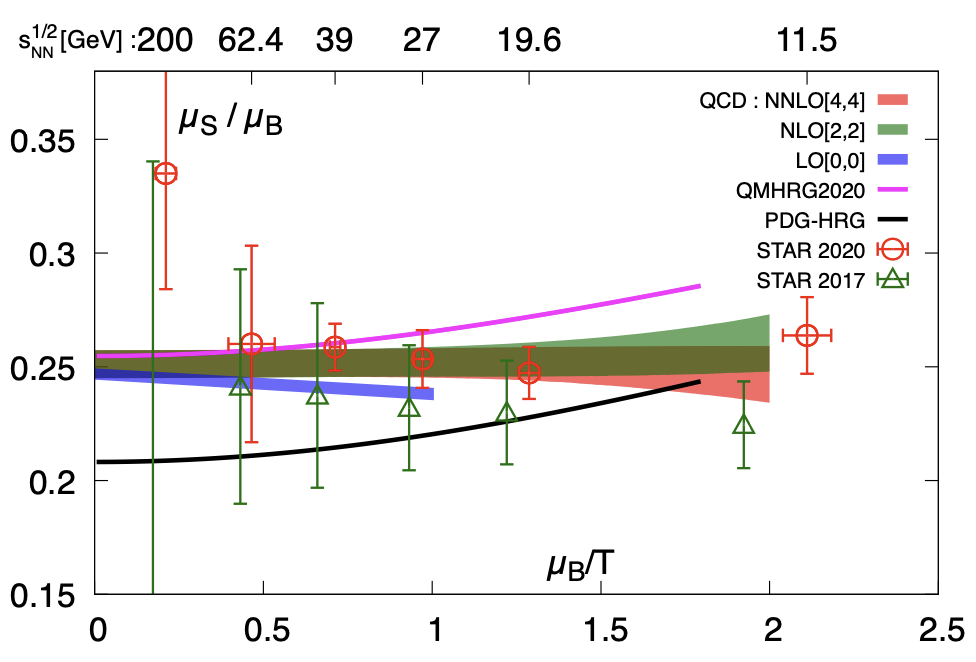}
  \includegraphics[width=0.31\linewidth]{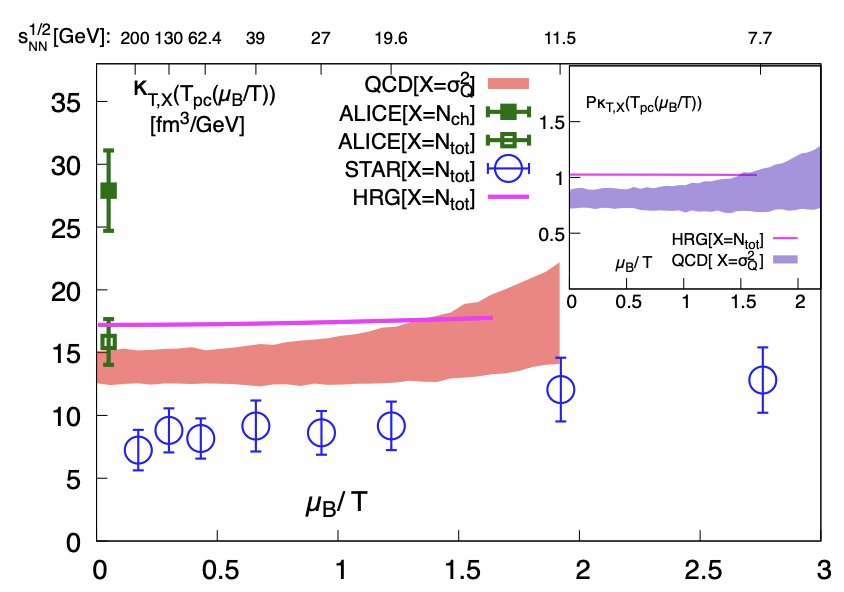}
  \includegraphics[width=0.31\linewidth]{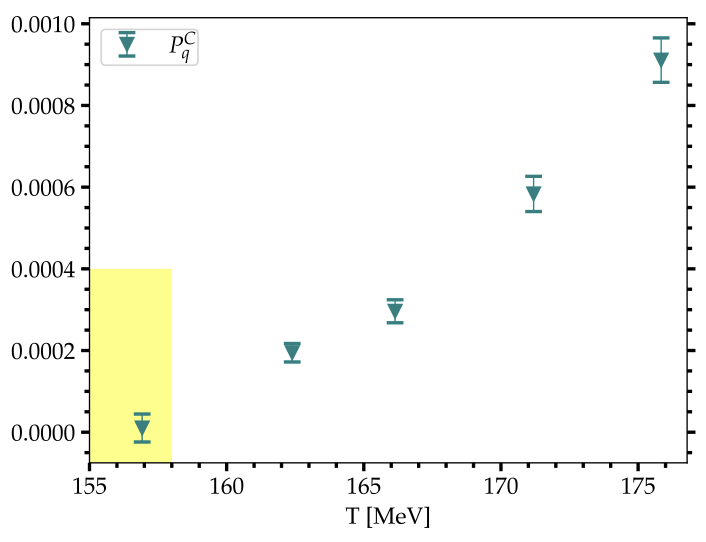}
\caption{Left: $\mu_S/\mu_B$ and strangeness correlations, confronting lattice QCD with STAR BES-II updates~\cite{Bollweg:2024epj}. 
Middle: generalized isothermal compressibility $\kappa_{T,\sigma_Q^2}$ from lattice QCD, compared with ALICE (charged-only and rescaled total-hadron), STAR, and HRG expectations~\cite{Clarke:2025gbj}. 
Right: charm-quark partial pressure across the crossover, illustrating the onset of quark-like charm degrees of freedom~\cite{Kaczmarek:2025dqt}.}
  \label{fig:bulk_thermodynamics}
\end{figure}

\paragraph{Strangeness chemical potential and correlations: lattice vs experiment.}
In heavy-ion collisions, approximate strangeness neutrality ($n_S=0$) and a fixed charge-to-baryon ratio ($n_Q/n_B\simeq 0.4$) constrain the thermodynamic trajectory in
$(T,\mu_B,\mu_Q,\mu_S)$ space.
Lattice QCD determines the corresponding chemical potentials (e.g.\ $\mu_S/\mu_B$) and conserved-charge correlations that can be mapped onto experimentally accessible cumulant ratios.
Along the pseudocritical line $T_{\rm pc}(\mu_B)$, HotQCD finds that isospin-limit sum-rule violations among second-order cumulants remain small across the RHIC BES range, implying a tight relation between $\chi_{11}^{BS}$ and $\chi_{11}^{QS}$ under strangeness neutrality~\cite{Bollweg:2024epj}.
A particularly clean quantity is $\chi_{11}^{BS}/\chi_2^S$, which agrees with preliminary STAR BES-II results for $\sqrt{s_{NN}}\gtrsim 39$~GeV, with increasing tension at lower beam energies~\cite{Bollweg:2024epj}.
The lattice prediction for $\mu_S/\mu_B$ can also be confronted directly with STAR extractions from strange-baryon yields, providing a quantitative lattice--experiment bridge for the chemical potentials themselves, see left panel of~\autoref{fig:bulk_thermodynamics}.

\paragraph{A generalized isothermal compressibility.}
In statistical mechanics the isothermal compressibility is defined as
$\kappa_T=-(1/V)(\partial V/\partial P)_{T,N}$, i.e.\ at fixed total particle number.
In QCD, however, the well-defined conserved quantities are the net charges $N_B$, $N_Q$, and $N_S$, while a conserved total multiplicity is not available from first principles.
Moreover, a naive replacement of $N$ by a net charge in $\kappa_T$ leads to a divergence as $\mu\!\to\!0$, since net densities vanish at $\mu_B=\mu_Q=\mu_S=0$~\cite{Clarke:2025gbj}.
Ref.~\cite{Clarke:2025gbj} resolves this by introducing a generalized compressibility $\kappa_{T,f_Q,r_Q,r_S}$ defined at fixed charge ratios
$r_Q\equiv N_Q/N_B$ and $r_S\equiv N_S/N_B$ and at fixed $f_Q$, where $f_Q$ can be chosen to ensure a finite $\mu\to0$ limit.
A particularly useful choice is to hold fixed the net-charge variance
$\sigma_Q^2 \equiv VT\,\frac{\partial^2 P}{\partial\mu_Q^2}$, leading to $\kappa_{T,\sigma_Q^2}$, which is expressible entirely in terms of $BQS$ cumulants and remains finite at $\mu=0$.
This construction is additionally motivated by the HRG observation that total hadron multiplicities are well approximated by net-charge fluctuations near freeze-out, making $\kappa_{T,\sigma_Q^2}$ a natural first-principles proxy for compressibility along heavy-ion trajectories~\cite{Clarke:2025gbj}. At $\mu_B=0$, the rescaled ALICE point agrees with the lattice result within errors, while the charged-only ALICE estimate lies substantially higher, as shown in the middle panel of~\autoref{fig:bulk_thermodynamics}.

\paragraph{Charm degrees of freedom.}
At temperatures above the crossover, charm contributes non-negligibly to the pressure and provides an additional handle on deconfinement.
Recent lattice calculations of generalized charm susceptibilities show that below $T_{\rm pc}$ the data are described by an HRG only if one includes additional charmed hadrons beyond the established PDG list (a ``QM-HRG'' extension), whereas a PDG-only HRG underestimates the charm-sector thermodynamics~\cite{Kaczmarek:2025dqt}.
Just above the crossover the HRG description breaks down abruptly, and a decomposition of the charm pressure into partial pressures of charmed baryons, charmed mesons, and charm quarks shows that the charm-quark partial pressure becomes nonzero above $T_{\rm pc}$, while hadron-like contributions can remain dominant up to $T\simeq 176$~MeV~\cite{Kaczmarek:2025dqt}. 
As illustrated in the right panel of~\autoref{fig:bulk_thermodynamics}, this points to a gradual emergence of quark-like charm degrees of freedom rather than an abrupt disappearance of open-charm hadron-like excitations at the crossover.
This motivates a consistent inclusion of charm in high-$T$ EoS modeling and in lattice--phenomenology interfaces extending above the crossover.

\section{Thermodynamics in external conditions}
\label{sec:external}
External conditions such as strong magnetic fields, isospin chemical potential, rotation, and spin polarization provide controlled deformations of QCD that
are both theoretically rich and phenomenologically motivated (e.g.\ magnetic fields in non-central heavy-ion collisions).

\begin{figure}[!htp]
  \centering
    \includegraphics[width=0.31\linewidth]{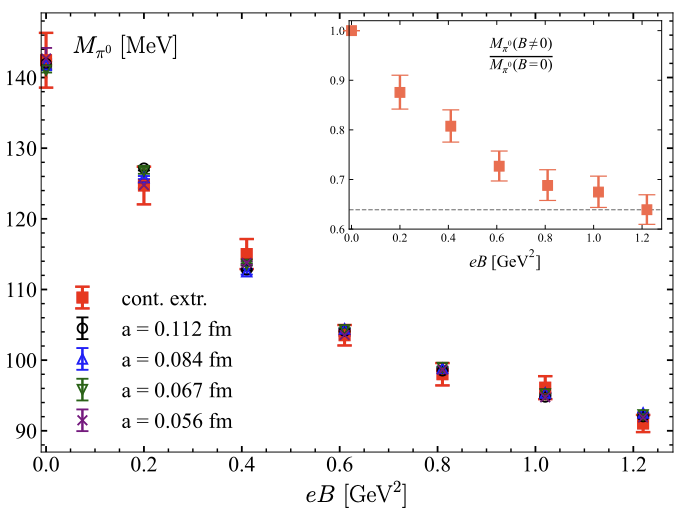}
  \includegraphics[width=0.31\linewidth]{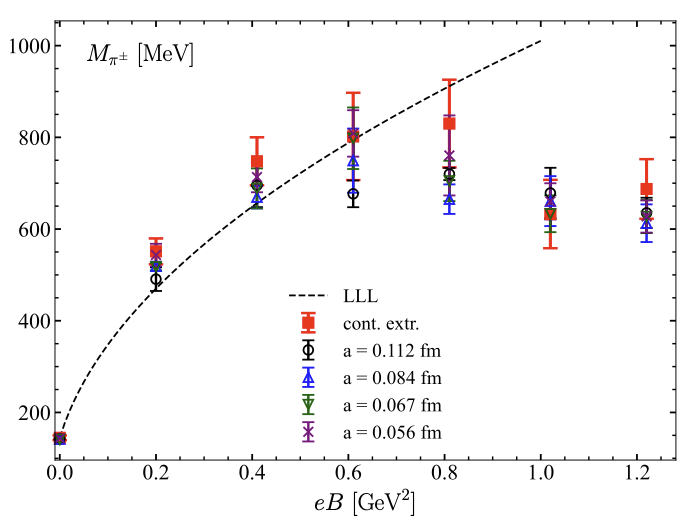}
  \includegraphics[width=0.31\linewidth]{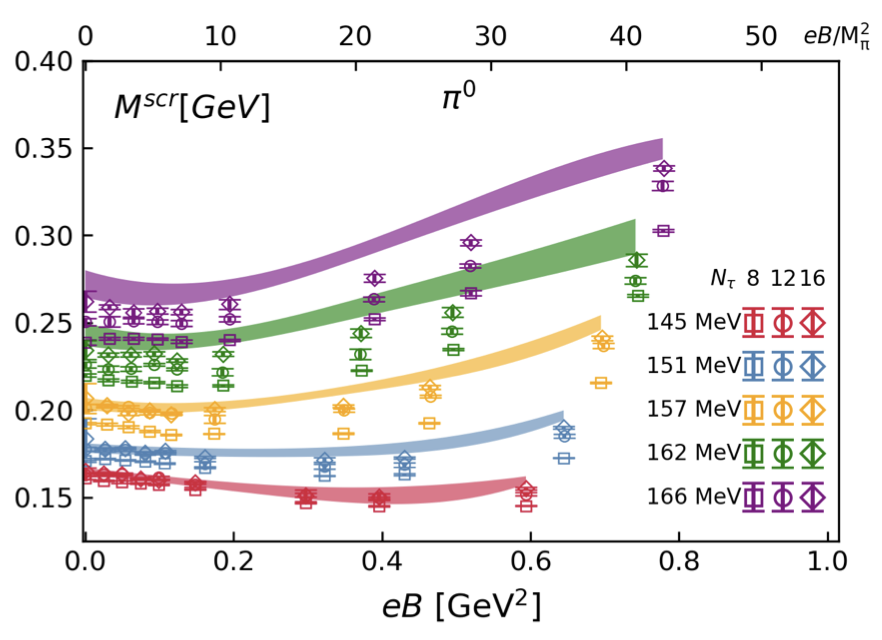}
\caption{Pions in external magnetic fields.
Left: continuum-extrapolated $\pi^0$ mass at $T\simeq0$ as a function of $eB$~\cite{Ding:2026qzu}.
Middle: Energy of $\pi^\pm$ at $T\simeq0$, showing an LLL-like rise at small $eB$ followed by saturation/turnover at larger fields~\cite{Ding:2026qzu}.
Right: continuum-estimated $\pi^0$ screening mass near the crossover; the bands denote the continuum estimates and illustrate its non-monotonic $eB$ dependence in the thermomagnetic medium~\cite{Ding:2025pbu}.}
  \label{fig:pion_extB}
\end{figure}

\paragraph{Pions in strong magnetic fields.}
In the vacuum, pions provide a benchmark for QCD in strong magnetic fields because they are the lightest (pseudo-)Goldstone modes and are directly sensitive to Landau quantization and chiral dynamics.
A recent $(2\!+\!1)$-flavor HISQ study at the physical point provides continuum-extrapolated results using four lattice spacings ($a\simeq0.112,\,0.084,\,0.067,\,0.056$ fm) and magnetic fields up to $eB\simeq1.22~\mathrm{GeV}^2$~\cite{Ding:2026qzu}.
It finds that the neutral-pion mass decreases steadily with $eB$, as shown in the left panel of~\autoref{fig:pion_extB}, while the charged pion exhibits a non-monotonic response: its lowest-Landau-level energy rises at small $eB$ in line with the point-like LLL expectation, but then saturates and slightly decreases at larger fields, signaling sizable internal-structure effects, as shown in the middle panel~\cite{Ding:2026qzu}.
A decomposition into valence and sea contributions shows that, at $T\simeq0$, the magnetic-field dependence of the pion masses is dominated by valence effects~\cite{Ding:2026qzu}.
Earlier simulations at a single lattice spacing with a heavier-than-physical pion mass ($M_\pi\simeq230$ MeV) already observed this turnover of the charged-pion energy and its LLL-like behavior at small fields, but without continuum control~\cite{Ding:2020hxw}.

At finite temperature near the crossover, screening masses provide a complementary long-distance probe.
Using Ward--Takahashi identities that relate pseudoscalar susceptibilities to chiral condensates, lattice calculations show that the $\pi^0$ screening mass closely tracks the competition between inverse magnetic catalysis and magnetic catalysis, leading to a characteristic non-monotonic dependence on $eB$ near $T_{\rm pc}$~\cite{Ding:2022tqn}.
The newer physical-mass study achieves continuum estimates with $N_\tau=8,12,16$ (aspect ratio $4$) for temperatures $T\simeq145$--$166$ MeV and fields up to $eB\sim0.8~\mathrm{GeV}^2$, establishing a controlled baseline for thermomagnetic screening phenomena; the resulting non-monotonic behavior of the $\pi^0$ screening mass is illustrated in the right panel of~\autoref{fig:pion_extB}~\cite{Ding:2025pbu,Thakkar:2025may}.

\paragraph{A QCD magnetometer and comparisons to ALICE.}
A major development is the proposal of a ``QCD magnetometer'' based on conserved-charge correlations in a magnetic background.
The central idea is that a magnetic field reshapes the phase space of charged hadrons (Landau quantization), which feeds directly into the leading-order
fluctuations and correlations $\chi^{XY}_{11}$; among them, the baryon--electric-charge correlator $\chi^{BQ}_{11}$ is found to be particularly sensitive.
At pseudo-critical temperatures, $\chi^{BQ}_{11}$ starts to increase rapidly once $eB\gtrsim 2 M_\pi^2$ and can reach an $\mathcal{O}(2)$ enhancement around
$eB\simeq 8 M_\pi^2$ at the physical point~\cite{Ding:2023bft,Ding:2025jfz,Ding:2025qzh,Ding:2025dvn}.
By contrast, a recent continuum-estimated study of the static-quark entropy in background magnetic fields finds no significant shift of the deconfinement-like pseudocritical temperature in the weak-field regime, with the entropy-based estimate remaining consistent with that from the total chiral susceptibility~\cite{Thakkar:2025lat}.

\begin{figure}[!htp]
  \centering
          \includegraphics[width=0.32\linewidth]{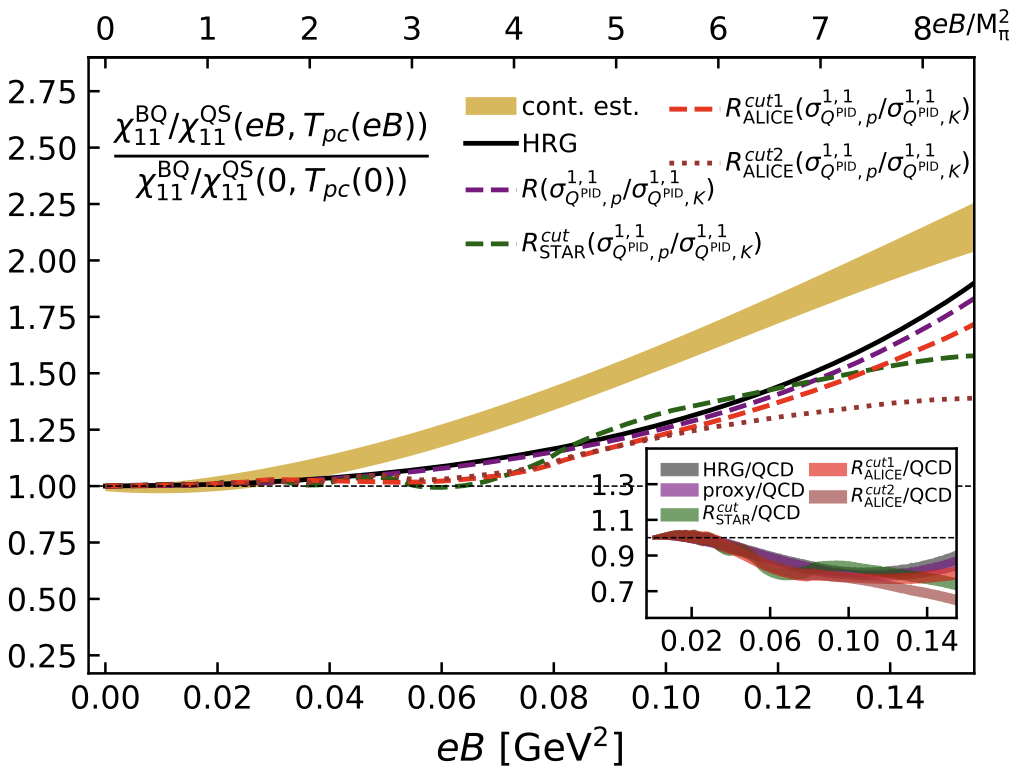}
    \includegraphics[width=0.32\linewidth]{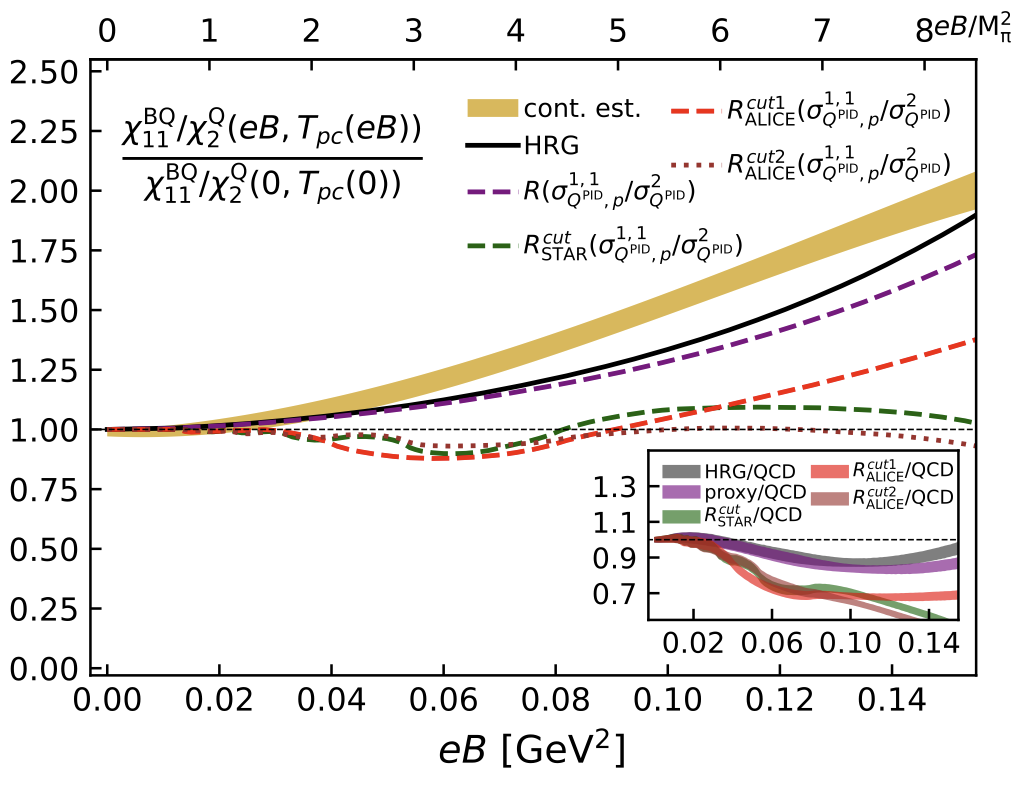}
 \includegraphics[width=0.32\linewidth]{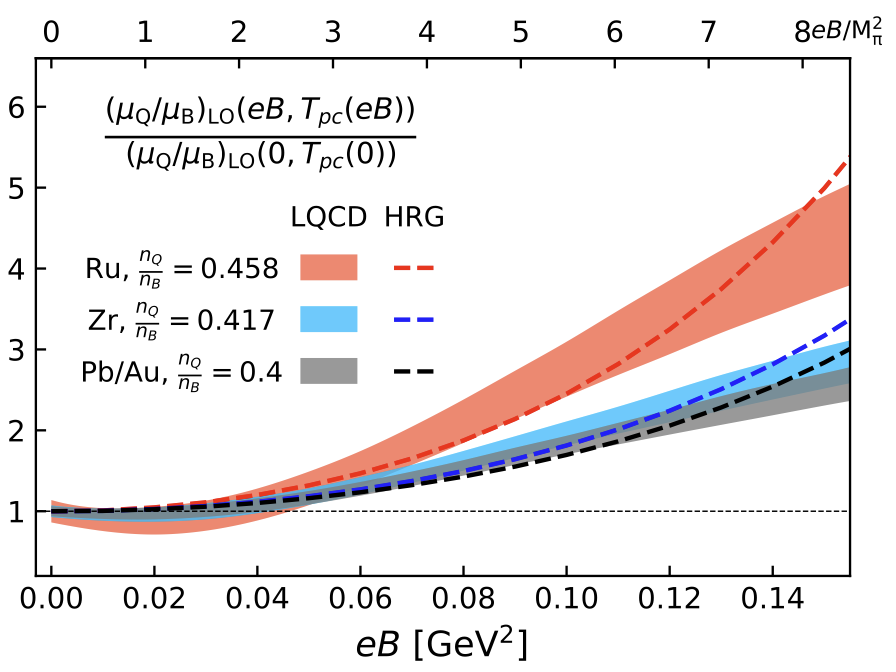}
\caption{Magnetometer observables in a magnetic background.
Left: Normalized double ratio $\chi_{11}^{BQ}/\chi_{11}^{QS}$ along the transition line, predicted to be more sensitive to $eB$ and less affected by common systematics~\cite{Ding:2025jfz}.
Middle: Normalized $\chi_{11}^{BQ}/\chi_2^{Q}$ along the transition line~\cite{Ding:2025jfz}.
Right: $\mu_Q/\mu_B$, illustrating both magnetic-field and system dependence~\cite{Ding:2023bft}.}
  \label{fig:magnetometer}
\end{figure}

To reduce trivial volume dependence and sharpen sensitivity to $eB$, it is useful to form normalized ratios such as
$\chi^{BQ}_{11}/\chi_2^{Q}$.
Lattice data exhibit the ordering
$\chi^{BQ}_{11}/\chi_{11}^{QS}>\chi^{BQ}_{11}/\chi_2^{Q}>\chi^{BQ}_{11}/\chi_2^{B}>\chi^{B}_{2}/\chi_2^{Q}$ with the double ratio $\chi^{BQ}_{11}/\chi^{QS}_{11}$ being the most sensitive to the magnetic field.
It reaches an enhancement of about $2.25$ at $eB\simeq 8 M_\pi^2$ along $T_{\rm pc}(eB)$~\cite{Ding:2025jfz}.
The first two of these ratios are shown in the left and middle panels of~\autoref{fig:magnetometer}.
Recent ALICE measurements of centrality-dependent conserved-charge correlations provide a first qualitative cross-check of  $\chi^{BQ}_{11}/\chi_2^{Q}$~\cite{ALICE:2025mkk} which follows the trend of lattice QCD data. Finally, the closely related quantity $\mu_Q/\mu_B$ is predicted to be sensitive not only to $eB$ but also to the collision system (e.g.\ isobars),
providing an additional handle complementary to correlation measurements~\cite{Ding:2023bft,Grefa:2026meq}.

\paragraph{Dense and magnetized equation of state.}
The combined effect of magnetic fields and baryon density on the EoS is of direct interest for both heavy-ion collisions and dense-matter applications.
At small $\mu_B$, the leading baryon-density dependence of the pressure in a magnetic background is encoded in
$c_2(B,T)\equiv \left.\frac{1}{2}\frac{\partial^2 (P/T^4)}{\partial \hat\mu_B^2}\right|_{\hat\mu_B=0}$,
which is directly related to baryon-number fluctuations and therefore provides a sensitive probe of how $B$ modifies the response of QCD matter.

Recent lattice studies have begun to map $c_2(B,T)$, and related leading-order EoS coefficients, using complementary strategies.
Astrakhantsev et al. extract the relevant expansion coefficients from simulations at imaginary chemical potential in $(2+1)$-flavor staggered QCD with physical quark masses, using $\mu_u=\mu_d=\mu$ and $\mu_s=0$ and taking the continuum limit from $N_\tau=8,10,12$ at $eB=0.3,0.6,1.2~\mathrm{GeV}^2$; the results indicate a substantial magnetic enhancement of the leading baryon-density response~\cite{Astrakhantsev:2024mat}.
A complementary $(2+1)$-flavor HISQ study at the physical point imposes strangeness neutrality together with the baseline condition $n_Q/n_B=0.4$, and obtains continuum-estimated leading-order EoS coefficients on $N_\tau=8,12$ lattices for $145\lesssim T\lesssim165~\mathrm{MeV}$ and $eB\lesssim0.8~\mathrm{GeV}^2$, including the LO pressure coefficient $P_2$ (equivalent to $c_2$ up to the chosen constraint)~\cite{Ding:2025nyh,Kumar:2025ikm}.
Building on their earlier low-density Taylor-expansion study, the recent work of Marques Valois et al. computes the dense and magnetized EoS from simulations at imaginary $\mu_B$ with $2+1+1$ flavors under heavy-ion constraints, finding strong $B$-induced modifications, especially near the crossover, while also emphasizing that a continuum extrapolation remains necessary for precision phenomenology~\cite{Borsanyi:2023buy,MarquesValois:2025nzo}.

Overall, these first-principles results consistently indicate that magnetic fields can substantially enhance the leading baryon-density response of the EoS near the crossover.
The quantitative differences among the various studies can be understood largely in terms of the imposed constraints on $\mu_Q$ and $\mu_S$ (for example, $\mu_S=0$ versus strangeness neutrality), as well as the flavor content of the simulations ($(2+1)$ versus $(2+1+1)$).

\begin{figure}[!htp]
  \centering
  \includegraphics[width=\linewidth]{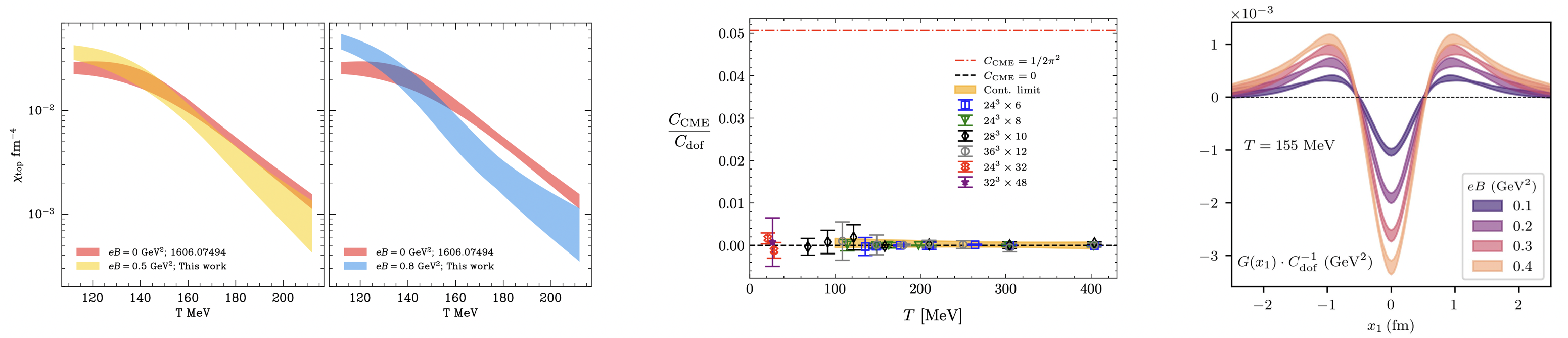}
\caption{Left: topological susceptibility in background magnetic fields. Middle: global (homogeneous) equilibrium CME indicator. Right: localized CME response in a non-uniform magnetic field. Plots (left to right) are taken from Refs.~\cite{Brandt:2024gso,Brandt:2024wlw,Brandt:2024fpc}.}
  \label{fig:cme_chitop_extB}
\end{figure}

\paragraph{Topology and anomaly-related observables in magnetic fields}
Continuum-extrapolated lattice calculations have recently provided the first quantitative mapping of the topological susceptibility $\chi_{\rm top}(T,eB)$ in $(2+1)$-flavor QCD at the physical point for fields up to $eB\simeq0.8~{\rm GeV}^2$ across the crossover region~\cite{Brandt:2024gso}. To control staggered cutoff effects associated with would-be zero modes, these studies employ gradient-flow definitions of $Q_{\rm top}$ together with an eigenvalue-reweighting strategy and emphasize the ratio
$R_\chi(B,T)=\chi_{\rm top}(B,T)/\chi_{\rm top}(0,T)$.
The main qualitative outcome, illustrated in the left panel of~\autoref{fig:cme_chitop_extB},  is that strong magnetic fields enhance $\chi_{\rm top}$ at low temperature, while the effect is gradually reduced and turns into a suppression around and above the crossover, i.e. $R_\chi<1$ at higher $T$~\cite{Brandt:2024gso}. This change of trend adds a new, anomaly-sensitive facet to inverse magnetic catalysis phenomena and has been discussed in connection with proposals of “axial inverse magnetic catalysis” near the transition~\cite{Wang:2021dcy}, motivating further continuum and chiral-limit studies.

Equilibrium studies of the chiral magnetic effect (CME) clarify the distinction between global and local responses. For homogeneous backgrounds, the equilibrium CME conductivity (the linear coefficient relating a uniform current to a uniform chiral imbalance and magnetic field) is found to vanish once gauge-invariant regularization and conserved lattice currents are implemented, as shown in the middle panel of~\autoref{fig:cme_chitop_extB}~\cite{Brandt:2024wlw}. In contrast, in non-uniform magnetic fields a nontrivial localized CME response emerges: the induced current density develops a spatial profile correlated with the field and changes sign such that its volume integral (the global current) remains zero, as illustrated in the right panel~\cite{Brandt:2024fpc}. 
Recent work has also initiated first-principles studies of the out-of-equilibrium CME response, including a Euclidean-time correlator approach on the lattice and Kubo-formula based estimates of the corresponding conductivity~\cite{Buividovich:2024bmu,Brandt:2025now}.

\paragraph{Isospin chemical potential, spin polarization, rotation and acceleration.}
At finite isospin chemical potential, QCD enters a pion-condensed BEC regime once the isospin density becomes energetically favored; in the convention commonly used in lattice studies, this occurs at $\mu_I \gtrsim m_\pi/2$, equivalently $\mu_Q \gtrsim m_\pi$ on the pure charge axis.
Recent lattice determinations of the EoS at nonzero $\mu_I$ provide a controlled laboratory for finite-density physics, including characteristic structures of the interaction measure in the BEC phase and a stiffening of the EoS reflected in an enhanced speed of sound~\cite{Brandt:2022hwy}.
Going beyond the pure-$\mu_I$ axis, a leading-order Taylor expansion around simulation points at $\mu_I\neq 0$ enables access to small $\mu_B$ and $\mu_S$ deformations and, as a first application, yields the pressure along the pure charge axis $\mu_Q$, conveniently visualized as a three-dimensional surface $p/T^4=p/T^4(T,\mu_Q)$ and compared to the direct $\mu_I$-axis results deep into the BEC regime~\cite{Brandt:2024dle,Brandt:2022hwy}. 
Complementary insight into the onset of pion condensation at finite isospin chemical potential has also been obtained from an unbiased exponential resummation of Lee--Yang zeros, yielding estimates compatible within uncertainties with the pion-condensation threshold~\cite{Mitra:2024czm}.

Two other externally controlled deformations are quark spin polarization and rotation.
For a finite quark spin potential $\mu_\Sigma$, studied via analytic continuation from imaginary $\mu_\Sigma$, pseudocritical temperatures extracted from the Polyakov-loop susceptibility and from the disconnected chiral susceptibility follow a quadratic dependence on $\mu_\Sigma$; the resulting positive curvature implies that real spin polarization lowers both $T_{pc}^{\rm conf}$ and $T_{pc}^{\chi}$~\cite{Braguta:2025ddq}.
For rotation, lattice simulations at imaginary angular velocity reveal a spatially inhomogeneous confinement--deconfinement structure in gluodynamics, with confined and deconfined regions separated by a radius-dependent boundary; analytic continuation to real rotation then suggests a deconfined core near the rotation axis and a confined periphery~\cite{Braguta:2023iyx}. Recently, acceleration was first studied in lattice gluodynamics and found to strongly soften the deconfinement transition under weak acceleration~\cite{Chernodub:2024wis}; this was later reformulated in Rindler spacetime, yielding qualitatively similar behavior together with a spatially inhomogeneous confinement--deconfinement structure, whose boundary approximately follows the Tolman--Ehrenfest law~\cite{Braguta:2026kuz}.

\section{Summary}

Lattice QCD has become a precision tool for QCD thermodynamics at $\mu_B=0$ and is steadily extending its reach toward the chiral limit, low-temperature finite density, and thermodynamics under external conditions. Recent work strengthens the crossover picture at the physical point, sharpens the use of chiral and Dirac-spectrum observables near the chiral limit, and improves constraints on the phase boundary at $\mu_B>0$. At the same time, studies in strong magnetic fields, at finite isospin density, and under spin polarization, rotation, and acceleration continue to broaden the thermodynamic map of QCD and its phenomenological connections.

\section*{Acknowledgements}
This work is supported partly by the National Natural Science Foundation of China under Grants No. 12325508, No. 12293064, and No. 12293060 as well as the National Key Research and Development Program of China under Contract No. 2022YFA1604900 and the Fundamental Research Funds for the Central Universities, Central China Normal University under Grants No. 30101250314 and No.30106250152. 

\bibliographystyle{JHEP.bst}

\bibliography{refs.bib}

\end{document}